\documentstyle[a4wide,11pt]{article}
\author{Wolfgang Priester, Carsten van de Bruck \\
Institut f\"ur Astrophysik \\
der Universit\"at Bonn \\
Auf dem H\"ugel 71 \\
53121 Bonn, Germany}

\title{On the Origin of Very Wide Ly$\alpha$-Absorption-Lines in
Quasar Spectra}

\begin{document}

\maketitle

\begin{abstract}
We present a new explanation for the very wide absorption features in quasar
spectra. In our model, a very wide absorption feature will originate, when the
line of sight crosses a bubble wall tangentially. We demonstrate this on the
quasar pair (2138-4427), (2139-4434). Both show two very wide absorption
lines in their spectra at the same redshift. The bubble wall model can
explain these observations in low density Friedmann-Lemaitre models with
spherical metric. It contradicts models with euclidean or hyperbolic metric.
\end{abstract}

\section{Introduction}
The Ly$\alpha$ absorption systems have been analysed with the complete
Friedmann equation (i.e. including the $\Lambda$-term) by using two approaches:
1) A bubble wall distribution of hydrogen absorbers and 2) a cloud model with
homogeneously distributed absorbers (Hoell, Liebscher, Priester (1994),
hereafter HLP). The observations of high-resolution spectra can be explained
in both cases by the same Friedmann-Lemaitre model ($\Omega_{0}=0.014$,
\mbox{$\lambda_{0}=\Lambda c^2 / 3 H_{0}^2 = 1.080$}). A simple explanation for
this agreement is the fact, that the observed line counts for the cloud model
are usually averaged over redshift intervals $dz=0.2$. This implies averaging
over more than twenty bubble walls. Thus, a homogeneous distribution of clouds
is compatible with clouds situated in the walls of a bubble structure.

The bubble wall model is based on the assumption that the void-structure
observed in the distribution of galaxies in our cosmological neighbourhood
($z<0.05$) is a universal phenomenon and is expanding with the Hubble flow
only.
This predicts a charateristic pattern in the statistical distribution of
absorption lines in the Ly$\alpha$-forest (i.e. single lines or close blends
($\Delta v\leq 300$ km/s) representing hydrogen-filaments in the walls either
inside galaxies or in intergalactic clouds). High-resolution spectra of
numerous quasars in the redshift range $1.8<z<4.5$ show this characteristic
pattern (see e.g. Pettini et al. (1990)).

The aim of this letter is to show that the very wide Ly$\alpha$-absorption
features can also be explained with a universal, spongelike bubble wall
distribution of hydrogen absorbers. We shall demonstrate this on the pair
of quasars (1) 2138-4427 ($z_{em}=3.17$), (2) 2139-4434 ($z_{em}=3.23$). They
are separated by 8 arcmin on the sky. Both show two very wide absorption lines
in their spectra at the same redshift ($z_{abs}=2.380$ (4110 \AA) and
$z_{abs}=2.853$ (4685 \AA)), see Francis and Hewett (1993). Our
interpretation of these observations is presented in 2.
Conclusions are presented in 3.

\section{The interpretation with the bubble wall model}
The fact, that the two spectra of the quasar pair show similar absorption
features at the same redshifts, leads to the conclusion, that there are two
large objects which cross both lines of sight. In a Einstein-de Sitter
(flat) universe ($\Lambda \equiv 0$) the diameters of these objects would
be about $6\cdot h_{0}^{-1}$Mpc ($H_{0}=h_{0}\cdot 100$km/(s$ \cdot$ Mpc)).
It was pointed out by Peacock (1993), that these objects can not be
explained in the standard model of structure formation (CDM, HDM or MDM).

An obvious explanation for this astonishing finding is provided by the bubble
wall approach. It is assumed that the wide features originate when the lines
of sight from the two quasars to the observer cut tangentially through the
wall system on opposite sides of a void. This is shown schematically in
Fig. 1 and Fig. 2. The spectra centered at the absorption maximum at
4110 {\AA} and 4685 {\AA} were taken from Fig. 4 of Francis and Hewett
(1993). The spectral resolution ($1.3$ \AA) is not large enough to show a
clear pattern. Furthermore, the spectrum (A) is contaminated by possible
Ly$\beta$ absorption (quasar (1) Ly$\beta$(emission) at 4278 {\AA} and
quasar (2) at 4340 \AA) and both could be disturbed by metal lines.

Coinciding very wide absorption features can be expected in close quasar
pairs if the separation of the lines of sight corresponds to the expected
average size of bubbles at a certain redshift. In the case of the quasar
pair here we even have this occurrence twice at two redshifts. Does the
separation at $z=2.380$ and $z=2.853$ agree with the expected size? We
shall see that this is the case sufficiently well for low density
Friedmann-Lemaitre models with spherical space metric. It contradicts,
however, models with euclidean or hyperbolic metric (details see below).
The separation $d(z)$ between the lines
of sight at redshift $z$ is given by
\begin{equation}
d(z) = \frac{\alpha D_{r}(z)}{1+z}
\end{equation}
with $\alpha=$angular separation between the two quasars and
$D_{r}(z)=$metric distance, given by
\begin{equation}
D_{r}(z)=R_{0}\cdot r(z)
\end{equation}
with
\begin{equation}
R_{0}=\frac{c}{H_{0}}\sqrt{\frac{k}{\Omega_{0}+\lambda_{0}-1}}
\end{equation}
and
\begin{eqnarray}
r(z)=\left \lbrace \begin{array}{r}
\sinh \chi (z) \\
\chi(z) \\
\sin \chi(z) \end{array}\right\rbrace \begin{array}{r}
 \\
$for k=$ \\
\\
\end{array}\left\lbrace \begin{array}{r}
-1\\
0\\
+1\\
\end{array}\right\rbrace.
\end{eqnarray}
The radial distance $\chi(z)$ along the line of sight is given by
\begin{equation}
\chi(z)=\sqrt{\frac{\Omega_{0}+\lambda_{0}-1}{k}}\int_{1}^{1+z}
\frac{d\zeta}{\sqrt{\Omega_{0}\zeta^{3}-(\Omega_{0}+\lambda_{0}-1)\zeta^{2}
+\lambda_{0}}}.
\end{equation}
(Note, that for $k=0$ the quantity $R_{0}\cdot\chi(z)$ remains finite!).
The metric distance for four models is shown in Fig. 3.
In the bubble wall approach it is assumed that the size of
the bubbles (or voids) expands with the Hubble flow only. For the comparison
with the size of the Harvard voids in our neighbourhood
we transform d(z) to the present-epoch $t_{0}$:
\begin{equation}
d_{0} = d(z)\cdot(1+z).
\end{equation}
In addition we calculate the typical average bubble size
$\overline{R_{0}\Delta \chi}$ at the present-epoch, given by
\begin{equation}
\overline{R_{0}\cdot\Delta\chi}=\frac{\overline{\Delta z(z)} \cdot c}{H(z)}.
\end{equation}
The spectral resolution (1.3 \AA) and the possible contamination by $Ly\beta$
absorptions prevent us from deriving the typical seperation $\Delta z(z)$
between the absorption lines from the given spectra with sufficient
reliability.
We therefore use the average $\overline{\Delta z(z)}$ derived by HLP based
on numerous quasar spectra. We have, however, attempted to count the
line numbers in the spectra of the two quasars, whenever
it appeared possible, in order to derive the $\Delta z(z)$-values directly.
They could be considered as approximate lower limits. In this
sense we shall use it for calculating the bubble size along the line of
sight $R(z)\Delta\chi$ directly as approximate lower limit. The values
for the present-epoch are then obtained by multiplication with ($1+z$).
The values of $\Delta z(z)$ and $\overline{\Delta z(z)}$ are given in Tab. 1.
\begin{table}
\caption{Redshift intervals $\Delta z(z)$ in the line pattern obtained
from the spectra (A) and (B) (2nd column) and the typical average
values $\overline{\Delta z(z)}$ obtained by the Friedmann
regression analysis (HLP 1994) (3rd column)}
\begin{center}
\begin{tabular}{|l|l|l|}
\hline
z   & $\Delta z(z)$ & $\overline{\Delta z(z)}$\\
\hline
2.380 & $0.47\cdot 10^{-2}$ & $0.67 \cdot 10^{-2}$\\
\hline
2.853 & $0.52\cdot10^{-2}$ & $0.63\cdot 10^{-2}$\\
\hline
\end{tabular}
\end{center}
\end{table}

Here we use four representative models given in Tab. 2. The calculated
separations $d_{0}$ are given in Tab. 3 for spectrum (A) ($z_{abs}=2.380$,
given in Fig. 1) and in Tab. 4 of spectrum (B) ($z_{abs}=2.853$,
given in Fig. 2).
\begin{table}
\caption{The parameters of the four models for comparison ($c/H_{0}$ in Gpc)}
\begin{center}
\begin{tabular}{|l|l|l|l|l|}
\hline
Model & $h_{0}$ & $\Omega_{0}$ & $\lambda_{0}$ & $c/H_{0}$\\
\hline
Einstein-de Sitter (EdeS) & 0.5 & 1.00 & 0.0 & 6.00\\
\hline
Ostriker-Steinhardt (O-St) & 0.65 & 0.35 & 0.65 & 4.62\\
\hline
Sandage-Tammann (Sa-Ta) & 0.5 & 0.05 & 0.0 & 6.00\\
\hline
Bonn-Potsdam (BN-P) & 0.9 & 0.014 & 1.08 & 3.34\\
\hline
\end{tabular}
\end{center}
\end{table}
One can see, that only the BN-P-model gives a sufficient agreement between the
present diameters $d_0$ and $d_{void}=\Delta z(0)\cdot c/H_{0}$, where
$\Delta z(0)=0.009 \pm 0.002$, see deLapparent, Geller and Huchra (1986).

The other three models, however, can not explain the wide absorption
features in the bubble wall interpretation. This is evident in particular
if we compare the bubble sizes along the line of sight $R_{0}\cdot
\Delta \chi$ or $\overline{R_{0}\cdot \Delta \chi}$ (column 3 and 4)
with the Harvard void size $d_{void}$ \mbox{(column 5)} in Tab. 3 and Tab. 4.
\begin{table}
\caption{Calculated values for the four representative models in the
case of spectrum (A) ($z_{abs}=2.380$)}
\begin{center}
\begin{tabular}{|l|l|l|l|l|}
\hline
Model & $d_{0}$ & $R_{0}\Delta \chi$ & $\overline{R_{0}\Delta \chi}$
& $d_{void}$ \\
      &   Mpc   &    Mpc   &    Mpc   &    Mpc  \\
\hline
E-dS & 13 & 4.5 & 6.4 & 54 $\pm$ 12 \\
\hline
O-St& 14 & 5.7 & 8.2 & 42 $\pm$ 10 \\
\hline
Sa-Ta & 21 & 8.0 & 11.4 & 54$\pm$12 \\
\hline
BN-P & 19 & 21.5 & 30.7 & 30$\pm$7 \\
\hline
\end{tabular}
\end{center}
\end{table}
\begin{table}
\caption{Calculated values for the four representative models in the
case of spectrum (B) ($z_{abs}=2.853$)}
\begin{center}
\begin{tabular}{|l|l|l|l|l|}
\hline
Model & $d_{0}$ & $R_{0}\Delta \chi$ & $\overline{R_{0}\Delta \chi}$
& $d_{void}$\\
      &   Mpc   &   Mpc   &  Mpc  &   Mpc \\
\hline
E-dS & 14 & 4.3 & 5.2 & 54$\pm$12 \\
\hline
O-St & 15 & 5.5 & 6.7 & 42$\pm$10 \\
\hline
Sa-Ta& 24 & 7.8 & 9.5 & 54$\pm$12 \\
\hline
BN-P & 22 & 24.8 & 30.1 & 30$\pm$7 \\
\hline
\end{tabular}
\end{center}
\end{table}

\section{Conclusions}
We have seen, that a very wide absorption feature in quasar spectra
can originate, when the line of sight cuts tangentially through a
bubble wall in a universe in which a bubble structure expands with
the Hubble flow. The wide Ly$\alpha$ features consist of blends of
many lines produced in hydrogen clouds. We have demonstrated this
on the quasar pair (2138-4427), (2139-4434). We find a sufficient
agreement in closed low density cosmological models, here the BN-P
model, derived by Hoell, Liebscher and Priester (1994). Models
with euclidean or hyperbolic metric fail to explain the wide
absorption lines with the bubble wall approach.

We note that the metal lines (e.g. $^{16}$O, see, for example
Turnshek et al. (1989)) corresponding to the wide Ly$\alpha$
features are usually small. This is due at first to their larger
atomic weight and second to the finding that only a small
portion of the hydrogen clouds are already sufficiently enriched
with elements produced by stellar evolution, for example in the
inner parts of galaxies.

\newpage

\noindent {\bf References}
\begin{description}{}\itemsep=0pt\parsep=0pt\parskip=0pt
\topsep=0pt\labelsep=0pt
\item deLapparent, V., Geller, M.J. , Huchra, J.P.: ApJ
{\bf 302}, L1 (1986)
\item Francis, P.J., Hewett, P.C.: Astron. J. {\bf 105},
1633 (1993)
\item Hoell, J., Liebscher, D.-E., Priester, W.:
Astron. Nachr. {\bf 315}, 89 (1994) (HLP)
\item Peacock, J.: Nature {\bf 364}, 103 (1993)
\item Pettini, M., Hunstead, R.W., Smith, L.J., Mar, D.P.:
MNRAS {\bf 246}, 545 (1990)
\item Turnshek, D.A., Wolfe, A.M., Lanzetta, K.M.,
Briggs, F.H., Cohen, R.D., Foltz, C.B., Smith, H.E., Wilkes, B.J.:
ApJ {\bf 344}, 567 (1989)
\end{description}

\newpage
\begin{center}
\underline{{\LARGE Figure Captions}}\\
\end{center}
{\bf Fig. 1 :} Spectrum (A) of quasar (1) and (2) centered at
4110 {\AA} ($z_{abs}=2.38$) taken from Francis and Hewett (1993).
Our interpretation of this observation is shown schematically. \\
{\bf Fig. 2 :} The same as in Fig. 1 for spectrum (B) of both
quasars centered at 4685 {\AA} ($z_{abs}=2.853$) \\
{\bf Fig. 3 : }The metric distance for the four representative
models (see Table 2).

\end{document}